\documentclass[preprint,11pt]{elsarticle}
\pdfoutput=1

\usepackage{amssymb}
\usepackage{pifont}
\newcommand{\cmark}{\ding{51}}%
\newcommand{\xmark}{\ding{55}}%
\usepackage[table]{xcolor}
\usepackage{color}
\usepackage{array}
\usepackage{nicematrix}
\usepackage[export]{adjustbox}
\usepackage{textcomp}
\usepackage{lineno}
\usepackage{csquotes}
\usepackage[parfill]{parskip}

 \definecolor{blush}{rgb}{0.87, 0.36, 0.51}

\usepackage{hyperref}
\hypersetup{
    colorlinks=true,
    linkcolor=blush,
    citecolor=blush,
    filecolor=blush,      
    urlcolor=cyan
    }

\makeatletter
\providecommand{\doi}[1]{%
  \begingroup
    \let\bibinfo\@secondoftwo
    \urlstyle{rm}%
    \href{http://dx.doi.org/#1}{%
      doi:\discretionary{}{}{}%
      \nolinkurl{#1}%
    }%
  \endgroup
}
\makeatother

\journal{Transportation Research Part C}

\begin{document}

\begin{frontmatter}

\title{Towards more reliable public transportation Wi-Fi Origin-Destination matrices: Modeling errors using synthetic noise and optical counts}


\author[inst1,inst3]{Léa Fabre}
\author[inst2]{Caroline Bayart\footnote{Formerly at: Laboratoire de Sciences Actuarielles et Financières, Université Lumière Lyon 1, France.}$^{,}$}
\author[inst4,inst3]{Alexandre Nicolas}
\author[inst1]{Patrick Bonnel}

\affiliation[inst1]{organization={Laboratoire Aménagement Economie Transports, Université Lumière Lyon 2},
            addressline={3 rue Maurice Audin}, 
            city={Vaulx-en-Velin},
            postcode={69120}, 
            country={France}}

\affiliation[inst3]{organization={Explain},
            addressline={36 bd des Canuts}, 
            city={Lyon},
            postcode={69004}, 
            country={France}}

\affiliation[inst2]{organization={Laboratoire CHROME, Nîmes Université},
            addressline={Rue du Dr G. Salan}, 
            city={Nîmes},
            postcode={30021}, 
            country={France}}

\affiliation[inst4]{organization={Universite Claude Bernard Lyon 1, CNRS, Institut Lumière Matière},
            addressline={10 Rue Ada Byron}, 
            city={Villeurbanne},
            postcode={F-69622}, 
            country={France}}

\begin{abstract}
To continuously monitor mobility flows aboard public transportation, low-cost data collection methods based on the passive detection of Wi-Fi signals are promising technological solutions, but they yield uncertain results. We assess the accuracy of these results in light of a three-month experimentation conducted aboard buses equipped with Wi-Fi sensors in a sizable French conurbation. We put forward a method to quantify the error between the stop-to-stop origin-destination (O-D) matrix produced by Wi-Fi data and the ground truth, when the (estimated and real) volumes per boarding and alighting are known. To do so, the error in the estimated matrix is modeled by random noise. Neither additive, nor multiplicative noise replicate the experimental results. Noise models that concentrate on the short O-D trips and/or the central stops better reflect the structure of the error. But only by introducing distinct uncertainties between the boarding stop and the alighting stop can we recover the asymmetry between the alighting and boarding errors, as well as the correct ratios between these aggregate errors and the O-D error. Thus, our findings give insight into the main sources of error in the Wi-Fi based reconstruction of O-D matrices. They also provide analysts with an automatic and reproducible way to control the quality of O-D matrices produced by Wi-Fi data, using (readily available) count data.
\end{abstract}




\end{frontmatter}





\section{Introduction}
\label{sec:intro}
The past decades have been marked by several transformations such as urban sprawling, demographic growth and emergence of new transport modes (self-service bicycles, car sharing...) that have had a well known impact on mobility behaviors \citep{deschaintres2019analyzing, axhausen2002observing}. More specific events such as COVID-19 and the restrictive measures implemented by the government also disrupted public transit ridership. Along with these changes, demand for travel has turned less and less regular and professionals face greater uncertainty when it comes to modeling them.

In order to embrace uncertainty and propose solutions adapted to the needs of public transport users, analysts need to capture temporal and spatial variability of individual mobility behaviors \citep{gore2019exploring} and produce quality Origin-Destination (O-D) matrices \citep{KRISHNAKUMARI202038}. O-D surveys provide a detailed characterization of individual mobility behavior on a defined perimeter, but are only conducted every so often and on a limited sample of the population. 

In recent years, the information technology (IT) sector has been growing at a rapid pace and now offers various opportunities for mobility data collection. With the spread of portable electronic devices (smartphones, laptops, smartwatches…) and the growing acceptance of innovations by users \citep{puMonitoringPublicTransit2021}, a new horizon for crowdsourced data collection has emerged. This kind of data are now commonly used in transport planning and intelligent transport systems \citep{vlachogiannis2023intersense}. This includes GPS data, movement sensors, telephony data, Bluetooth and Wi-Fi data. Wi-Fi data deserve special attention, as this technology enables a larger sample of trips to be recorded at a very low cost \citep{traunmuellerDigitalTracesModeling2017}. This totally passive \citep{nittiIABACUSWiFiBasedAutomatic2020} data collection method performs rather well, especially in its ability to provide information about entire trips (not only the number of boarding and alighting passengers at a stop) in a variety of modes, and can be used to build stop-to-stop O-D matrices on a regular basis.

In a previous work, the comparison of O-D matrices built from Wi-Fi data collected in buses with those from on-board surveys has shown promising results \citep{fabre2023potential}. Nevertheless, challenges are still ahead of us. While the error rate between the two matrices is relatively low overall, this result may hide a more complex reality (for example, many errors between some O-D pairs and fewer on others). Then, the evolution of several factors (acceptance of the technology, characteristics of the connected objects, regulations...) are likely to alter the availability of passive data. In order to make greater use of Wi-Fi data in the future and get access to accurate passenger flows, analysts need a method to evaluate their quality on a regular basis.

To address this challenge, the paper proposes to quantify the error observed in stop-to-stop O-D matrices reconstructed from Wi-Fi data collected on a bus network. O-D surveys are implemented every ten years on average, which is not frequent enough to be used as a benchmark. In contrast, optical counts are reliable, inexpensive, and continuously available, as operators equip vehicles with counting cells, but only record the number of passengers boarding and alighting \emph{at a given stop}. This paper aims to better appraise the quality of Wi-Fi-based O-D matrix estimation by investigating the relation between the error on the aggregate boarding or alighting volumes per stop (using optical counts as the baseline) and the error on the detailed O-D trip shares (using onboard surveys as the baseline).

\begin{figure}[ht]\vspace*{4pt}
\centerline{\includegraphics[width=\textwidth]{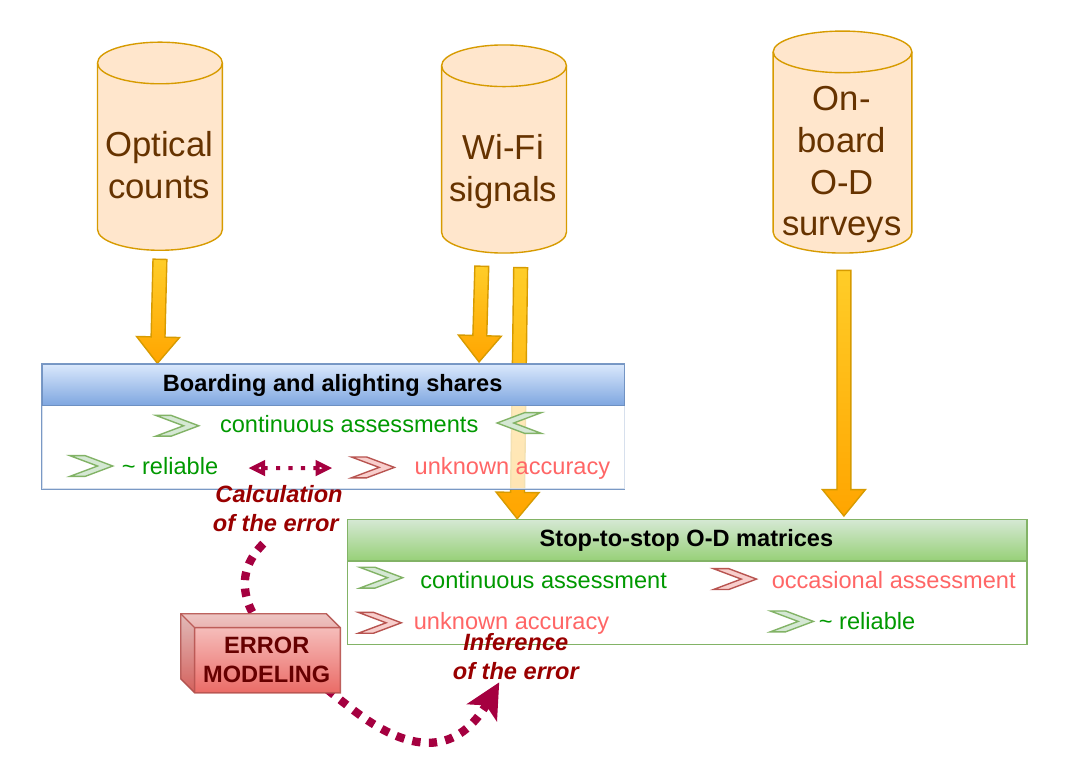}}
\caption{Workflow of the computation of boarding and alighting shares and of the construction of stop-and-stop O-D matrices based on different sources of data. The main contribution of the paper is the additional dark-red arrows that quantify the error on (continuously available) Wi-Fi based O-D matrices in light of continuously available data.}
\label{fig13}
\end{figure}

The originality of the method lies in the fact that we do not only compute the error, but we model its structure by leveraging ideas from stochastic processes; we are thus able to infer the main origin of the error and gain insight into the accuracy of the Wi-Fi sensors.
Our approach consists in modeling the error on the O-D matrix as a random matrix whose properties are to be ascertained and in focusing on the ratio between the error on boarding (or alighting) passengers shares per stop and the error on passengers shares per O-D trip. The structure of the noise matrix controls the expected ratio between these two indicators.
This leads to an automatic, reliable and easily transferable method to estimate the error on \emph{O-D pairs} when the error on \emph{volumes per boarding and alighting stops} is available. The logical architecture of the proposed methodology is synthesized in Figure~\ref{fig13}.

This paper briefly reviews the state of the art on emerging data sources for transportation and the assessment of their relevance (section 2). Section 3 details the model used to identify the link between the error on O-D pairs generated from Wi-Fi data and the errors on boarding and alighting shares using noise matrices. The Wi-Fi data and the methodology used to collect them and create O-D pairs are presented in section 4. Then, the paper presents how the error on O-D pairs behave with respect to the error on boarding or alighting passengers shares (section 5). Finally, we discuss the results and get insights for an enhanced use of Wi-Fi data in understanding mobility behavior (section 6).

\section{Related work}
\label{sec:related_work}

In recent years, the development of passive data collection and processing methods has attracted growing interest from both the scientific community and transport professionals \citep{BONNEL20181,DESCHAINTRES2023105079,bonnetain2019,su132011450,vlachogiannis2023intersense}. Whether it's GPS data, movement sensors, smart-card data, telephony and mobile applications' data or several kinds of sensors, like optical, infra-red, Wi-Fi and Bluetooth, each emerging data source has its own characteristics (Table \ref{table1}). Note that several of these methods target only a specific group of travelers (e.g., those who own a smartphone and use its Wi-Fi, or those who have a smart card, to the exclusion of fare evaders), which could introduce bias in some circumstances, unless the group is representative of the whole population.
\begin{table}[!ht]
\tiny
\centering
\caption{Advantages and drawbacks of different data sources.}
\rowcolors{2}{white!90!black!40}{white!70!black!40}
\makebox[\textwidth][c]{%
\begin{tabular}
{p{2cm}|m{1.3cm}m{0.8cm}m{1.7cm}m{1.7cm}m{1.7cm}m{1.4cm}m{1.3cm}}
\hline
 &Surveys&GPS&Movement sensors&Smart-card&Telephony&Sensors (optical, infra-red)&Wi-Fi/ Bluetooth \\
\hline
Passive & \hfil \textcolor{orange!80!black}{\xmark} & \hfil \textcolor{orange!80!black}{\xmark} & \hfil \textcolor{orange!80!black}{\xmark} & \hfil \textcolor{green!60!black}{\cmark} & \hfil \textcolor{green!60!black}{\cmark} & \hfil \textcolor{green!60!black}{\cmark} & \hfil \textcolor{green!60!black}{\cmark} \\
Multi-modal & \hfil \textcolor{green!60!black}{\cmark} & \hfil \textcolor{green!60!black}{\cmark} & \hfil \textcolor{green!60!black}{\cmark} & \hfil \textcolor{orange!80!black}{\xmark} & \hfil \textcolor{green!60!black}{\cmark} & \hfil \textcolor{green!60!black}{\cmark} & \hfil \textcolor{green!60!black}{\cmark} \\

Continuous & \hfil \textcolor{orange!80!black}{\xmark} & \hfil \textcolor{orange!80!black}{\xmark} & \hfil \textcolor{orange!80!black}{\xmark} & \hfil \textcolor{green!60!black}{\cmark} & \hfil \textcolor{green!60!black}{\cmark} & \hfil \textcolor{green!60!black}{\cmark} & \hfil \textcolor{green!60!black}{\cmark}\\

Transport oriented & \hfil \textcolor{green!60!black}{\cmark} & \hfil \textcolor{green!60!black}{\cmark} & \hfil \textcolor{green!60!black}{\cmark} & \hfil \textcolor{orange!80!black}{\xmark} & \hfil \textcolor{orange!80!black}{\xmark} & \hfil \textcolor{green!60!black}{\cmark} & \hfil \textcolor{green!60!black}{\cmark} \\

Origin + Destination & \hfil \textcolor{green!60!black}{\cmark} & \hfil \textcolor{green!60!black}{\cmark} & \hfil \textcolor{green!60!black}{\cmark} & \hfil \textcolor{orange!80!black}{\xmark} & \hfil \textcolor{green!60!black}{\cmark} & \hfil \textcolor{green!60!black}{\cmark} & \hfil \textcolor{green!60!black}{\cmark} \\

Link between O and D & \hfil \textcolor{green!60!black}{\cmark} & \hfil \textcolor{green!60!black}{\cmark} & \hfil \textcolor{green!60!black}{\cmark} & \hfil \textcolor{orange!80!black}{\xmark} & \hfil \textcolor{green!60!black}{\cmark} & \hfil \textcolor{orange!80!black}{\xmark} & \hfil \textcolor{green!60!black}{\cmark} \\

Fraud/Non-validation & \hfil \textcolor{green!60!black}{\cmark} & \hfil \textcolor{green!60!black}{\cmark} & \hfil \textcolor{green!60!black}{\cmark} & \hfil \textcolor{orange!80!black}{\xmark} & \hfil \textcolor{green!60!black}{\cmark} & \hfil \textcolor{green!60!black}{\cmark} & \hfil \textcolor{green!60!black}{\cmark} \\

Exhaustive & \hfil \textcolor{orange!80!black}{\xmark} & \hfil \textcolor{orange!80!black}{\xmark} & \hfil \textcolor{orange!80!black}{\xmark} & \hfil \textcolor{orange!80!black}{\xmark}/\textcolor{green!60!black}{\cmark} & \hfil \textcolor{orange!80!black}{\xmark}/\textcolor{green!60!black}{\cmark} & \hfil \textcolor{orange!80!black}{\xmark}/\textcolor{green!60!black}{\cmark} & \hfil \textcolor{orange!80!black}{\xmark} \\

\hline
\end{tabular}
}
\label{table1}
\end{table}

Among them, smart-card data and optical sensors give a good picture of mobility behaviors. They are both easy to get, since they are passively collected too, as in \citet{deschaintres2019analyzing} or \citet{wang2021two}, but present significant disadvantages. Optical counts are informative about the number of people getting on and off at a given stop, but they do not give the link between the origin and the destination of the trip. Smart-card data provide ticket validation information upon boarding only \citep{HUSSAIN2021103044}, even if researchers have developed algorithms to retrieve the alighting stops thanks to trip chains \citep{munizaga2012estimation} or more recently with probabilistic models \citep{cheng2021probabilistic}. So, these two data sources cannot be used to construct O-D matrices. Telephony data could give information about the origin and the destination of a trip \citep{paipuri2020estimating,alexander2015origin}, but they are the property of telecom operators and are very expensive to get \citep{calabrese2013understanding}; however, they allow one to target a variety of transport modes, and not only public transportation.
Wi-Fi and Bluetooth data seem particularly promising, since they are collected passively and continuously. Most of all, they provide the link between the origin of a trip and its destination. Wi-Fi or Bluetooth sensors are embedded in vehicles or positioned along the roads and can detect connected objects, without identifying them personally. Devices can be uniquely identified using Media Access Control (MAC) ID under short-range communication protocols \citep{sharma2020analysis, jiEstimatingBusLoads2017, fukudaEstimationParatransitPassenger2017}.  By using information from a GPS, it is then possible to build O-D matrices. Some experimentation stated that Wi-Fi is a better candidate than Bluetooth to gather individual trips in buses because of a quicker detection and a greater field of action \citep{hidayatWiFiScannerTechnologies2018,kurkcuEstimatingPedestrianDensities2017,paradedaBusPassengerCounts2019}. This study is therefore restricted to Wi-Fi sensors.

The use of Wi-Fi data in transport planning tends to democratize, and there are already a few studies working with them to represent trips along a transport network \citep{dunlapEstimationOriginDestination2016,mishalaniUseMobileDevice2016,fukudaEstimationParatransitPassenger2017,haakegaard2018statistical,jiEstimatingBusLoads2017,grgurevic2022overview}. Among the most recent ones,\citep{nittiIABACUSWiFiBasedAutomatic2020,afshariIntelligentTrafficManagement2019a,hidayat2020estimating} focus on the number of passengers boarding or alighting at a stop or on bus loads. \citet{puMonitoringPublicTransit2021} estimate O-D matrices by using Wi-Fi sensors to collect data in buses and machine learning to retrieve stop-to-stop O-D matrices. The results are compared with manual counts, and the authors mentioned the good quality of Wi-Fi data despite the non-exhaustive representation of passenger mobility on the network. \citet{algomaiah2022utilizing} go a little bit further in the use of Wi-Fi data to understand mobility behaviors by optimizing the detection radius of Wi-Fi sensors and most of all by inferring transfer activities to passengers of the bus. Finally, \citet{chang2023online} developed an online ridership estimation algorithm using Wi-Fi sensors as entry data. Comparison with manual counts and with other algorithms confirmed the "feasibility of estimating online ridership through passive Wi-Fi sensing". Similarly, \citet{fabre2023potential} as well as \citet{wang2022bus} use machine-learning and other algorithms to build O-D matrices from Wi-Fi data and prove the reliability of the latter with a comparison to ground truth data. Results are promising and encourage the use of these kinds of sensors. This Wi-Fi based technology could be used to probe travel behaviors with a fine temporal granularity, out of the reach of conventional surveys, but at present the uncertainty about their true accuracy would obscure the distinction between the part of the variations over time due to actual mobility changes and that due to detection and reconstruction error.

Most studies compare Wi-Fi O-D matrices and ground truth matrices through the lens of performance measures. In most of the previously cited studies, performance is measured with global metrics (R$^2$, RMSE, MAPE) to give an order of magnitude of how close to the reality O-D matrices from Wi-Fi data are. Only O-D flows are compared, and no attention is paid to the structure of the error. Claiming that studying O-D matrix structural information is necessary to validate the relevance of the methods used to compute O-D matrices, some studies tend to assess structural similarities between estimated and reference O-D matrices. \citet{djukic2013reliability} introduced a new quality measure that takes into account structural similarities and  \citet{BEHARA2020513} proposed the "normalized Levenshtein distance for OD matrices (NLOD)". They both demonstrated the method robustness in giving a metric for structural comparison of O-D matrices.
On the other hand, some studies, mostly using traffic count data to generate O-D matrices, succeeded in bounding the error between estimated and reference O-D matrices depending on the number and positions of sensors used \citep{bierlaire2002total,gan2005traffic}. Most of these works are limited to the use of metrics to estimate how close a measurement is to reality. They provide interesting avenues of research for estimating the error between an O-D matrix calculated from passive data and the ground truth, but do not identify or model the error on the estimated matrix. Analysts therefore lack insight into the uncertainty of these estimates.

Accordingly, there is a gap in the state of the art when it comes to real-time evaluation of the error in Wi-Fi-based O-D matrix estimation, in the absence of the ground truth O-D matrix. 

\section{Data \& Methods}
\label{sec:methods}

\subsection{Optical counts}
\label{sub:optical}
To gauge the boarding and alighting volumes at every stop, we resorted to optical counts, namely, Automatic Passenger Counts systems (APC) obtained from sensors placed on-board buses. They reportedly provide very "accurate estimates of the vehicle on-board loads" \citep{RONCOLI2023103963}. Ref.~\cite{olivo2019empirical} reported a $95\%$ and $97\%$ accuracy for boarding and alighting passenger counts respectively when including a variance of plus or minus one passenger. Other studies have also reported good matching between APC and ground truth \citep{APC_Barabino}. The drawbacks of these sensors, besides the fact that it does only give the number of passenger boarding or alighting at a bus stop without providing information on the O-D trips, is the small number of vehicles equipped ($10\%$ to $20\%$ in France) \citep{RONCOLI2023103963}. Nonetheless,  they fulfill their role in our study, since we concentrate on buses equipped with both Wi-Fi sensors and optical sensors, and the latter are necessary complements to Wi-Fi data to get information on O-D trips (\emph{see below}).

\subsection{Wi-Fi signals}
\label{sub:data_WiFi}

\subsubsection*{Principle}

\begin{figure}[ht]\vspace*{4pt}
\centerline{\includegraphics[height=4cm]{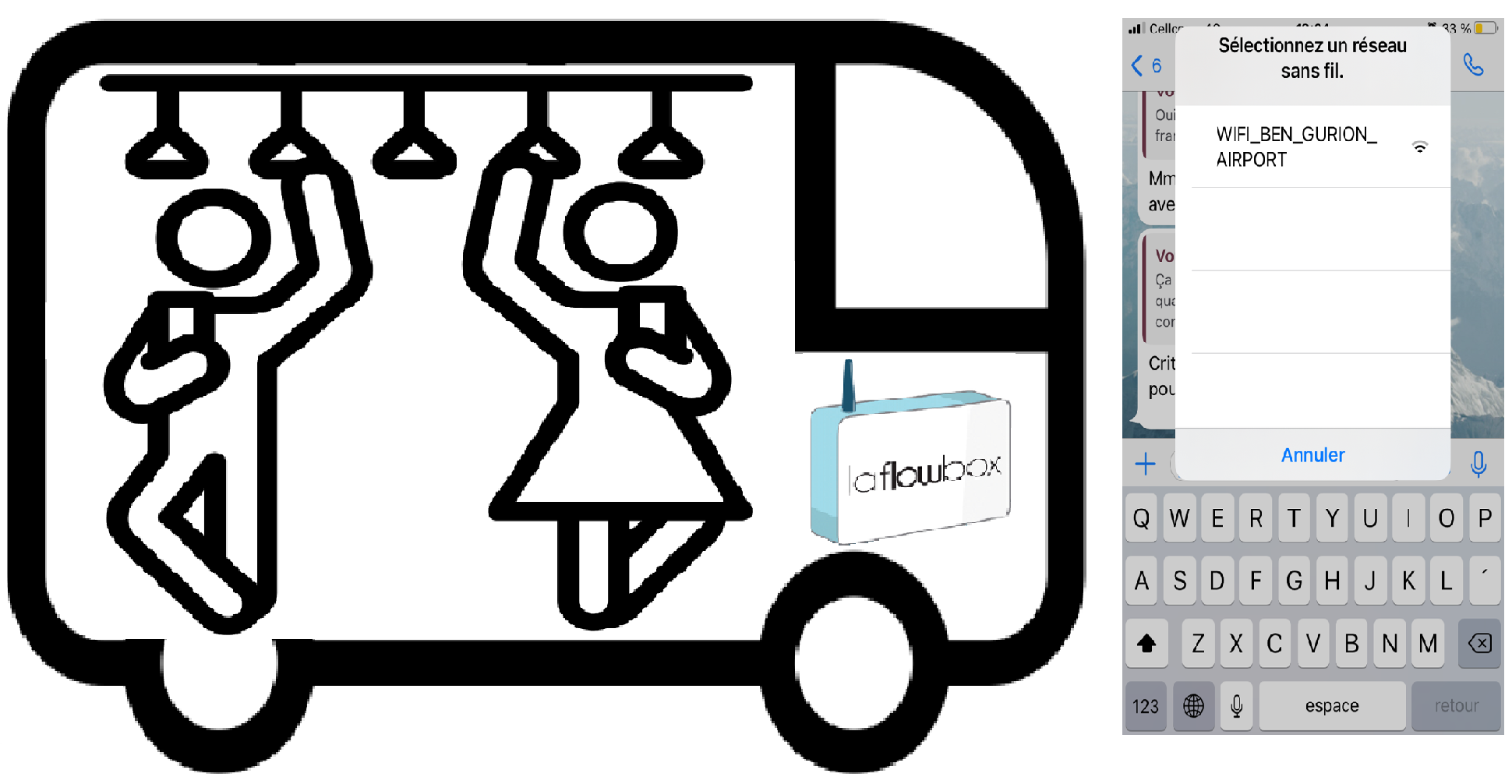}}
\caption{\textit{Laflowbox} device aboard buses and an example of a op-up that appears on connected objects when they search for a Wi-Fi connection.}
\label{fig1}
\end{figure}

The sensor used in this study is called \textit{Laflowbox}. It measures mobility using an electromagnetic wave sensor, which detects the signals emitted by connected objects which are seeking a Wi-Fi point to connect to (Figure~\ref{fig1}) (smartphone, smartwatch…). The sensor does not interact with the connected objects, which makes the data collection completely passive. The Wi-Fi file collected with this set-up is made of (encrypted) MAC addresses, the latter being considered personal data by the General Data Protection Regulation (GDPR) (Article 4(1)). However, this data collection complies with the French laws, as the MAC addresses are pseudonymized \cite{cnilLoiInformatiqueLibertes}. Moreover, posters displayed in the bus warn passengers that some sensors are active. \textit{Laflowbox} is also capable of recording the precise coordinates of connected objects every second, thanks to a GPS module. GPS and Wi-Fi data are then linked by a common timestamp.

\subsubsection*{Context of the data collection campaign}

Data were gathered in Rouen, a large city in the Normandy region (North of France). With a bit more than 111,000 inhabitants Rouen is the largest city of the Métropole Rouen Normandie (70 cities, 494,000 inhabitants) and a rather important economic hub at the national scale, mainly thanks to its seaport, the fifth largest in France. \textit{Laflowbox} sensors were placed in buses of the TEOR (Transport Est Ouest Rouennais) network, which gathers 30\% of the traffic flow and observes a constant increase since its implementation. This attractive network is composed of 4 lines, covering 65 stops and serving 7 other cities. The first line, T1, holds 15 stops and is about 8 km long. The T2, is much longer with 30 stops for over 30 km and the T3 holds 27 stops and is about 14 km long. These three lines share a common section of 10 stops in the city center. 

From November 1, 2019, to January 30, 2020 two boxes were placed on buses driving along lines T1, T2, and T3 alternatively. 
These two buses were also equipped with cameras for optical counts to facilitate comparison of the results of the two data sources. 

\subsubsection*{Data processing}
Considering the optical counts as a reference, the Wi-Fi sensors might be able to detect up to approximately 35\% of the passengers. However, among the millions of signals detected, many are spurious signals that need to be removed. 

Accordingly, the raw data were preprocessed to build the O-D matrix, following the methodology detailed in \cite{fabre2023potential,fabre2024estimating}. First, a large number of variables are computed to enrich the database (related to the signal, the device...). Then, we have to distinguish the signals emitted by real bus passengers from the interfering signals (people waiting or passing near the bus stops but not in the vehicle, people cycling on the road, people in their cars behind the bus...), which are removed from the database.
To filter the signals and to retain only those emitted by actual bus passengers, a clustering algorithm is used, as described in \citet{fabre2023potential,fabre2024estimating}. It uses the K-means algorithm to separate the signals detected by the Wi-Fi sensor into several groups, based on the principal components derived from numerous features related to the number of detections of the same connected object, the homogeneity of the detections of the same object, and so on. Each group formed by the K-means has unique characteristics. According to them, some clusters are defined as passenger signals and others as unwanted signals. The latter are eliminated. Finally, three stop-to-stop O-D matrices (one for each line of the network, ) are constructed from the passenger signal clusters, representing the number of passengers detected for each O-D pair of the corresponding line. The Wi-Fi estimated O-D matrices are aggregated over the entire data collection period (approximately three months) in order to obtain enough observations. The temporal aggregation is also a way to represent an average day over the data collection period to be comparable to O-D surveys, which are known to represent trips over a network on an average day.

\begin{figure}[ht]\vspace*{4pt}
\centerline{\includegraphics[height=6cm]{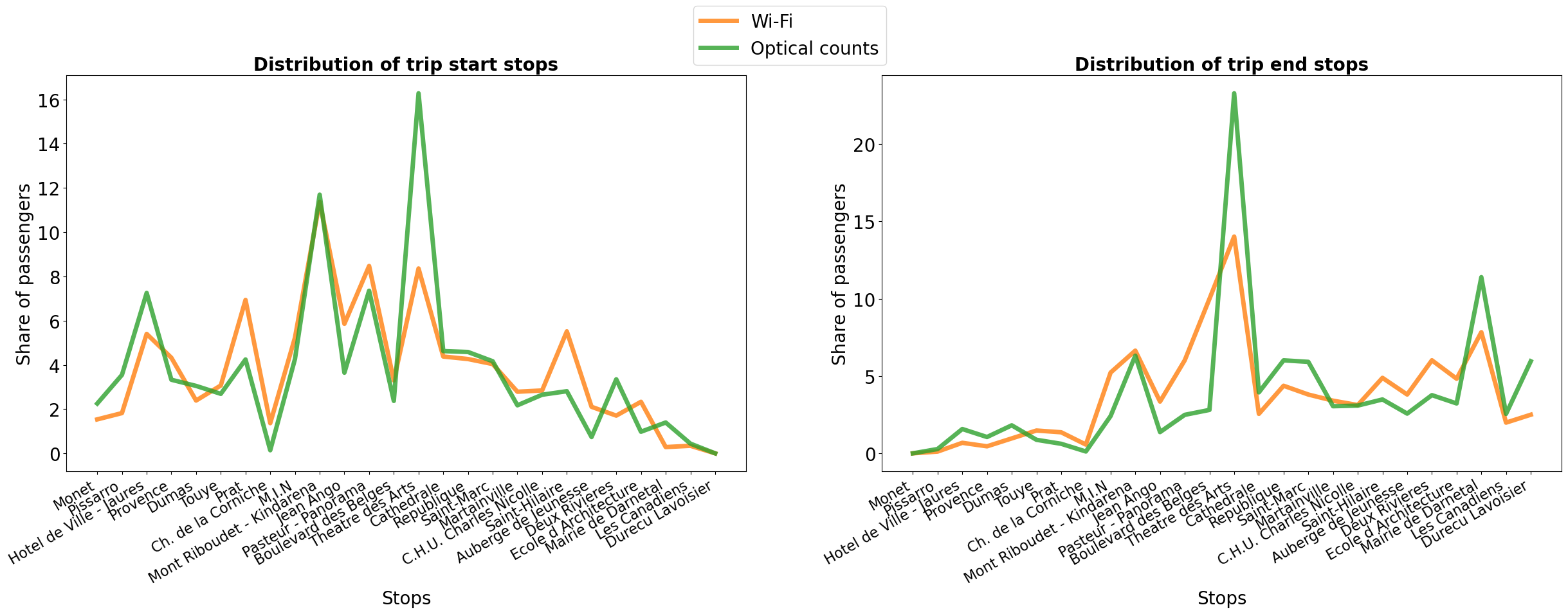}}
\caption{Distribution of trip start stops – T3 direction 1 – 2019/2020, in \%; (b) Distribution of trip end stops – T3 direction 1 – 2019/2020, in \%.}
\label{fig2}
\end{figure}

Once processed, the trips estimated using Wi-Fi sensors are in good agreement with those from other ground truth data sources (O-D surveys and optical counts), with $R^2$ values going up to 0.97 in the best case. To illustrate this result, figure \ref{fig2} shows the distribution of trip start (resp. end) points for Wi-Fi data and optical counts (for line T3, direction 1).

\subsection{O-D surveys}
\label{sub:data_surveys}

The last available stop-to-stop O-D survey in this territory was conducted onboard the T1, T2 and T3 lines, on 11$^{\mathrm{th}}$, October 2018, approximately one year before Wi-Fi data collection. Considering the average periodicity of O-D surveys (ten years), we can reasonably assume that the results of this survey can be used as a benchmark to assess the quality of the Wi-Fi matrices described above. 

\begin{figure}[ht]\vspace*{4pt}
\centering
\includegraphics[draft=False,width=\textwidth]{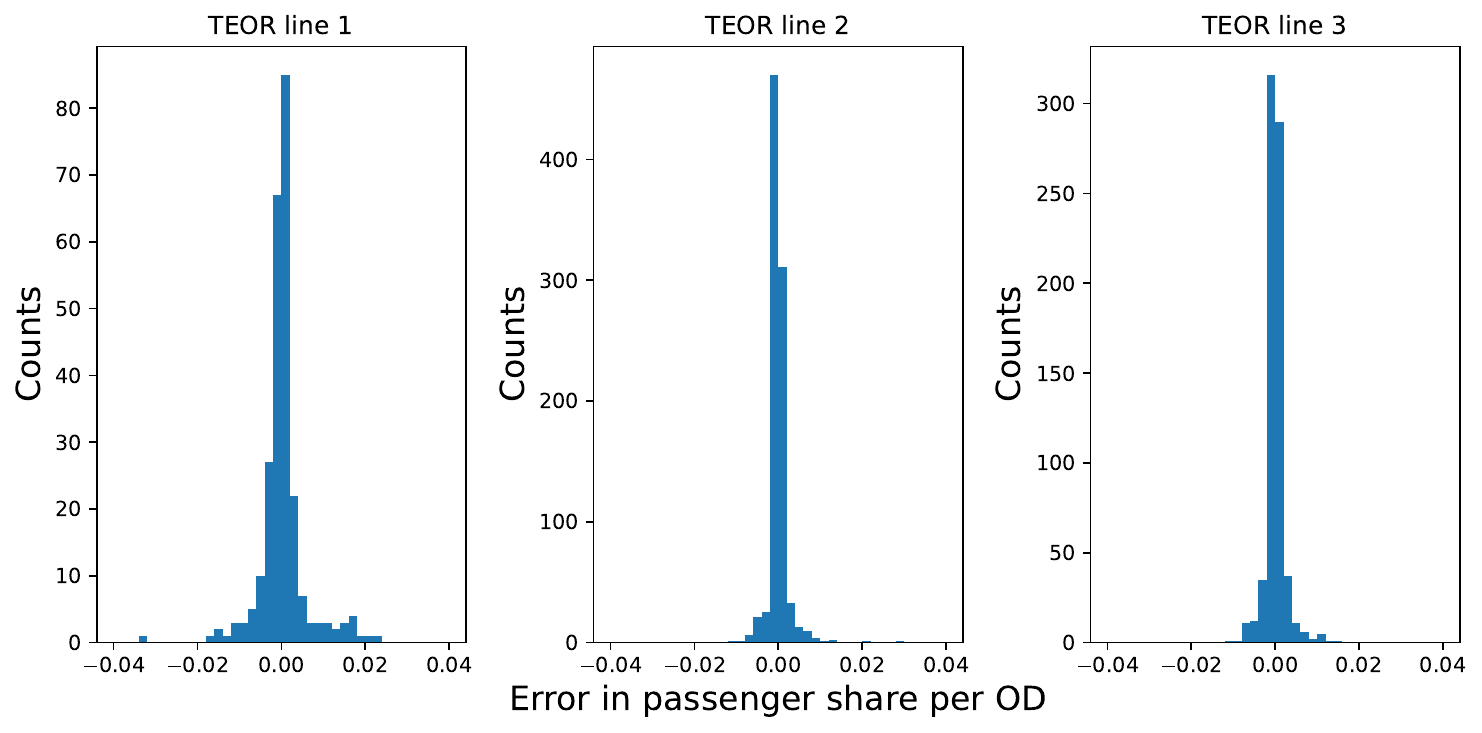}
\caption{Histograms of the errors in passenger share per O-D, calculated as the difference between the survey results and the \textit{Laflowbox} estimates  (in \%) per O-D pair, for the 3 TEOR lines: T1, T2 and T3.}
\label{fig7}
\end{figure}

The shares per O-D pair obtained from Wi-Fi data match reasonably well those with those from the survey, with $98\%$ of the observations for bus lines T2 and T3 having a difference between Wi-Fi and O-D survey shares of less than $0.005\%$ (it is $91\%$ for T1); the percentages are given as a function of the global volume. The magnitude of these differences are shown on Figure~\ref{fig7}.

\subsection{Errors on O-D trips and on alighting or boarding volumes}
\label{sub:methods_notations}

To refer to boarding/alighting stops, we will use the transparent notations $i$ (for in) and $o$ (for out). The stops will be labelled from $i=1$ (or equivalently $o=1$) to $i=N$ ($o=N$), where the number of stops $N$ depends on the transit line.

Let $[T_{io}]$ and $[\hat{T}_{io}]$ be the reference and estimated matrices of O-D shares:  $T_{io}$ is the share of travelers going from O to D, among all travelers, so that $\sum_{i=1}^{N}\sum_{o=1}^{N} T_{io} = 1$, and similarly for $[\hat{T}_{io}]$. We assume that the surveys give a good enough reflection of this ground truth; they are thus used as reference ($[T_{io}]$), while the estimated matrix $[\hat{T}_{io}]$ is based on Wi-Fi data   

We denote by $[\delta T_{io}]=[\hat{T}_{io}-T_{io}]$ the error matrix and we use the shorthand $T_{i\bullet}=\sum_{o}T_{io}$ for the reference boarding share at stop $i$ (i.e., the total number of passengers boarding at stop $i$ over the total number of passengers) and $T_{\bullet o}=\sum_{i}T_{io}$ for the reference alighting share at stop $o$. $\hat{T}_{i\bullet}$ and 
$\hat{T}_{\bullet o}$ are the estimated counterparts of $T_{i\bullet}$ and $T_{\bullet o}$, and $\delta T_{i\bullet}$ and $\delta T_{\bullet o}$ the differences between reference and estimate.

In our efforts to quantify the error $[\delta T_{io}]$, we will calculate the root-mean square (RMS) error on O-D shares (hereafter called O-D error),

\begin{equation}
Err \equiv \sqrt{ \frac{1}{N^2} \sum_{i=1}^{N} \sum_{o=1}^{N} \left(\delta T_{io}\right)^{2}},
\label{eq:Err}
\end{equation}
but also the RMS errors on boarding and alighting shares (hereafter called boarding and alighting errors, respectively)

\begin{equation}
Err^{(in)}\equiv  \sqrt{ \frac{1}{N} \sum_{i=1}^{N} \left( \delta T_{i\bullet}\right)^{2}}
\label{eq:Err_in}
\end{equation}

\begin{equation}
Err^{(out)}\equiv  \sqrt{ \frac{1}{N} \sum_{o=1}^{N} \left( \delta T_{\bullet o}\right)^{2}}
\label{eq:Err_out}
\end{equation}
and the error \emph{ratios} defined by
\begin{equation}
err^{(in)}\equiv \frac{Err^{(in)}}{Err} \text{ and } err^{(out)}\equiv \frac{Err^{(out)}}{Err}.
\label{eq:err_in_out}
\end{equation}

\subsection{Synthetic random noise matrices}
\label{sub:methods_noise}
To model the structure of the estimation error (using the 2018 O-D survey as reference), we will describe the error matrix $[\delta T_{io}]$ as a random noise matrix using minimal models for the noise.  Generally speaking, noise encompasses everything that cannot be described or modeled deterministically.

The simplest noise model assumes that the errors are uncorrelated between O-D shares, independent of the O-D share $T_{io}$, and uniformly distributed (in the same range for all $io$), i.e.
\begin{equation}
\delta T_{io} \sim \xi_{io}^{\mathrm{(add)}} = \xi_{io} - Z^{\mathrm{(add)}}[\xi]
\label{eq:add}
\end{equation}
where the $\sim$ operator means that two random variables follow the same distribution and the normalization function $Z^{\mathrm{(add)}}[\xi]=N^{-2}\,\sum_{i=1}^{N} \sum_{o=1}^{N} \xi_{io}$ is introduced to conserve the sum of shares, $\sum_{i=1}^{N}\sum_{o=1}^{N} \hat{T}_{io} = 1$. 
Here and henceforth, the random variables $\xi_{io}$ will always be independent and identically distributed (i.i.d), following a uniform distribution on the segment $[-\sigma,\sigma]$, with $\sigma>0$. 
We will refer to $ \xi_{io}^{\mathrm{(add)}}$ as \emph{additive} noise.  

A second model relies on the idea that the share error could be all the larger as the O-D share is large, which we describe with \emph{multiplicative} noise  
\begin{equation}
\delta T_{io} \sim \xi_{io}^{\mathrm{(mult)}} = \xi_{io}\, T_{io} - Z^{\mathrm{(mult)}}[\xi],
\label{eq:mult}
\end{equation}
where $Z^{\mathrm{(mult)}}[\xi]= N^{-2}\, \sum_{i=1}^{N} \sum_{o=1}^{N} \xi_{io} T_{io}$

Note that, while the foregoing synthetic noises conserve the sum of shares, they may entail that some shares become negative ($\hat{T}_{io}<0$) (or larger than 1), which is impossible in reality. This can be remedied by bounding all random realizations of $\xi_{io}$ that give rise to negative shares  $\hat{T}_{io}<0$. Then, in these cases, $\hat{T}_{io}$ is set to zero, and the normalization function $Z[\xi]$ is adjusted accordingly. When these safeguard operations are performed, we will say that the noise is \emph{clamped}; in practice, this clamping has no incidence on the results that we will present, except for the noise models concentrating on short and/or central O-D pairs (see below).

Anticipating on the remarks we will make below, we also introduce a noise model which concentrates the error on short O-D trips, i.e.,
\begin{equation}
\delta T_{io} \sim \xi_{io}^{\mathrm{(short)}} = \xi_{io} \, \Theta(2-|i-o|) - Z^{\mathrm{(short)}}[\xi],
\label{eq:short}
\end{equation}
where $\Theta$ is the Heaviside function ($\Theta(x)=1$ if $x\geqslant 0$ and $0$ otherwise) and 
 $Z^{\mathrm{(short)}}[\xi]=N^{-2}\,\sum_{i=1}^{N} \sum_{o=1}^{N} \xi_{io}\, \Theta(2-|i-o|)$,
as well as a noise model which concentrates the error on geographically central stops, which happen to correspond to labels $i$ and $o$ between $N/8$ and $3N/8$ (other choices can be made, without altering our conclusions),
\begin{equation}
\delta T_{io} \sim \xi_{io}^{\mathrm{(central)}} = \xi_{io} \cdot  \mathbf{1}_c(i) \cdot  \mathbf{1}_c(o) - Z^{\mathrm{(central)}}[\xi],
\label{eq:central}
\end{equation}
where $\mathbf{1}_c(x)=\Theta(x-\frac{N}{8})\cdot \Theta(\frac{3N}{8}-x)$ is the indicator function of the segment $[\frac{N}{8},\frac{3N}{8}]$, and 
 $Z^{\mathrm{(short)}}[\xi]=N^{-2}\,\sum_{i=1}^{N} \sum_{o=1}^{N} \xi_{io} \cdot  \mathbf{1}_c(i) \cdot  \mathbf{1}_c(o)$.

Finally, we define noise structures that operate on the boarding stop and the alighting stop separately. More precisely, the following noises mimic uncertainties regarding the boarding stop and the alighting stop, respectively,
\begin{equation}
\delta T_{io} \sim \xi_{io}^{\mathrm{(in)}} = \xi_{i} - Z^{\mathrm{(in)}}[\xi]
\label{eq:in-noise}
\end{equation}
\begin{equation}
\delta T_{io} \sim \xi_{io}^{\mathrm{(out)}} = \xi_{o} - Z^{\mathrm{(out)}}[\xi],
\label{eq:out-noise}
\end{equation}
where the $\xi_{i} \in [-\sigma, \sigma]$ and $\xi_{o} \in [-\sigma',\sigma']$ are independent and uniformly distributed, $Z^{\mathrm{(in)}}[\xi]= N^{-1} \sum_{i=1}^{N} \xi_{i}$, and $Z^{\mathrm{(out)}}[\xi]= N^{-1} \sum_{o=1}^{N} \xi_{o}$.

Of course, all these noise models can be combined additively, yielding \emph{inter alia} a \emph{combined additive and multiplicative} noise
\begin{equation}
\delta T_{io} \sim \xi_{io}^{\mathrm{(add-mult)}} = \xi_{io} + \xi^{\prime}_{io}\, T_{io} - Z^{\mathrm{(add)}}[\xi] - Z^{\mathrm{(mult)}}[\xi^{\prime}],
\label{eq:add-mult}
\end{equation}
where $\xi^{\prime}_{io}\sim \mathcal{N}(0,{\sigma'}^2)$.

\subsection{Tests using a synthetic O-D matrix}
\label{sub:synth-OD}
\begin{figure}[ht]\vspace*{4pt}
\centerline{\includegraphics[height=10cm]{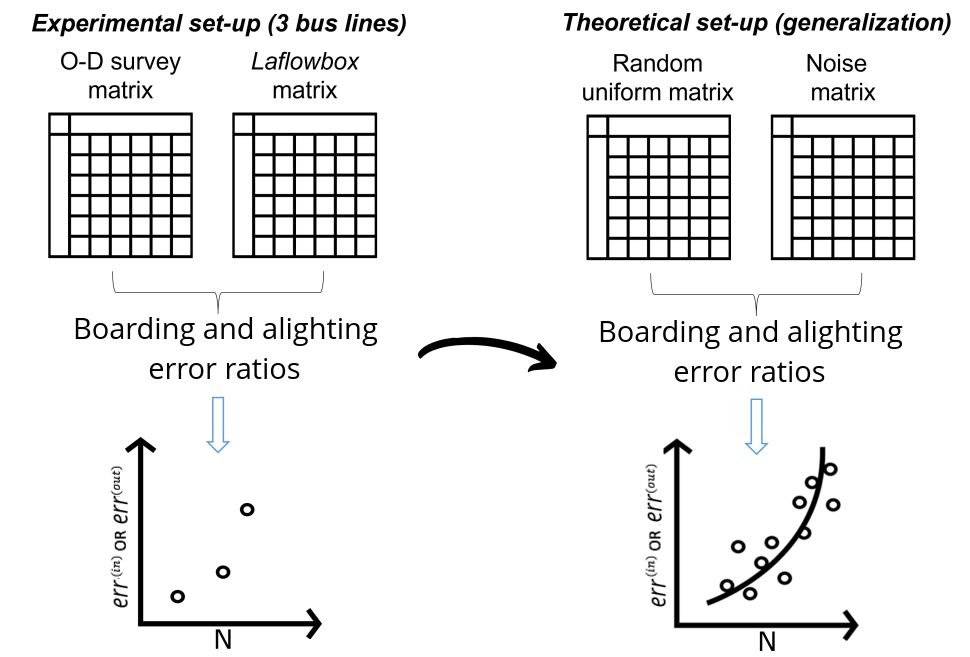}}
\caption{General methodology used to relate the errors on the O-D trip shares to the errors on boarding or alighting passenger shares. Comparison of the reference matrix (or its synthetic analogue) to the estimated matrix allows us to measure the mean O-D error and the mean boarding and alighting errors, whose dependence on the number of stops $N$ is plotted in a graph. }
\label{fig4}
\end{figure}

Inferring a suitable noise structure directly from the data is difficult, because of the limited data at our disposal: our experimental data only cover three bus lines and the O-D survey was performed on a specific day only. Taking inspiration from the creation of artificial noisy data used in machine learning algorithms assessment, as in \citet{khayrallah2018impact} and \citet{xu2017zipporah}, we resort to an intermediate step relying on synthetic O-D matrices. Figure~\ref{fig4} summarizes the process.

The first step was to generate a matrix that will play the role the reference O-D matrix, but for an arbitrary number of stops $N$. For that purpose, we simply fill the $N\times N$ matrix with random scalar entries drawn from a uniform distribution over $[0,1]$ and eventually rescale these random terms so that they sum to unity. Adding a noise model for the error $[\delta T_{io}]$ yields a (synthetic) analogue of the estimated O-D matrix. On this basis, the trends of the O-D error, boarding error, and alighting error as a function of the noise model can be determined for any number of stops $N$ (typically, between 1 and 100), and ultimately compared to our somewhat limited empirical data. This extension of the possible numbers of spots will facilitate regression analysis.

\section{Results}
\label{sec:results}

\subsection{Empirical observations on the structure of the estimation error}
\label{sub:obs}
\begin{figure}[ht]\vspace*{4pt}
\centering
\includegraphics[draft=False,width=\textwidth]{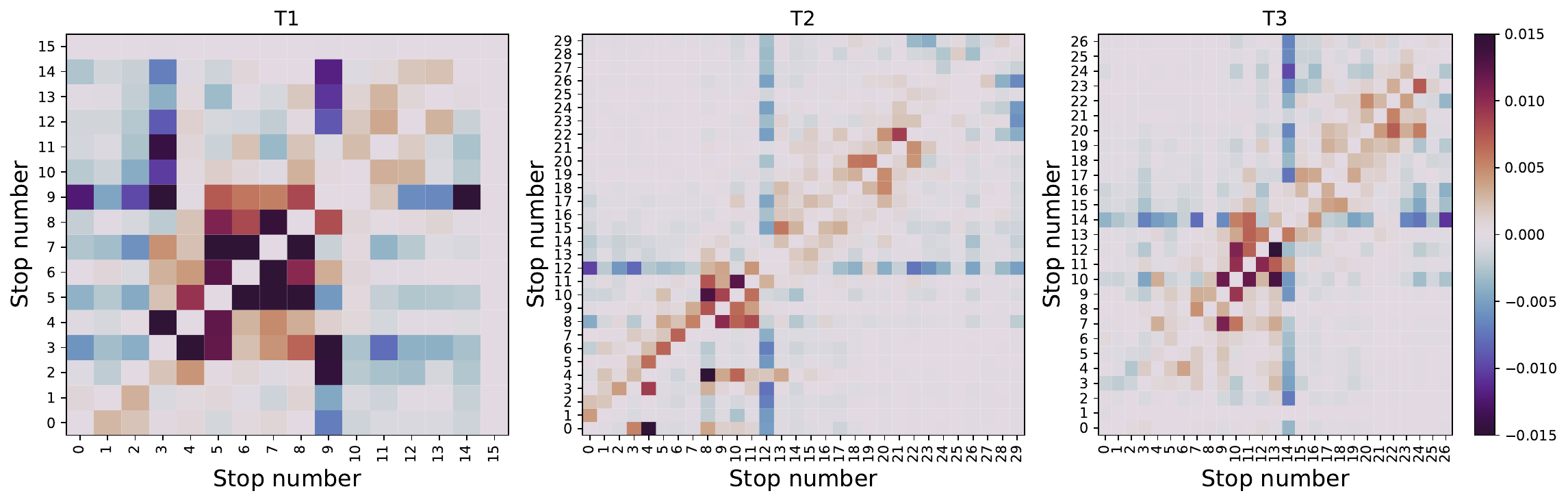}
\caption{Differences $\delta T_{io}$ of O-D pair shares (in \% of the total) for the three TEOR lines: T1, T2 and T3. Red and blue denote large positive and negative deviations, respectively;  the color scale saturates at the minimum and maximum values that are shown.}
\label{fig12}
\end{figure}

Figure~\ref{fig12} presents the O-D structure of the estimation error for the three bus lines under study, T1, T2 and T3, i.e., the difference $\delta T_{io}$ between the Wi-Fi-based estimate of the share $\hat{T}_{io}$ of passengers boarding at $i$ and alighting at $o$, and the reference represented by the survey (assimilated to the `ground truth').
Let us now comment on the most salient features,  which point to biases in the estimation.

Firstly, the error per O-D share seems to be greater for short O-Ds (along the diagonal) than for larger ones, and it tends to be positive: red hues can consistently be observed along the diagonal of the error matrix, on both sides, in Figure~\ref{fig12}, pointing to overestimated traffic shares on these O-D.
This may seem counterintuitive. Indeed, one can rightly argue that it is difficult to track the signals emitted by a connected object that does not travel long enough on the bus. In this case, those emitting at low frequencies are very likely not to be detected; on the contrary, it is easier for the sensors to detect several signals with a regular gap when the trip duration is long. However, irregular, infrequent, or poorly detected Wi-Fi emissions also result in the impression that trips are shorter than they actually are, notably producing a glut of very short trips.

Secondly, the error is also larger (more positive) in central locations. This also makes sense.
Indeed, at downtown stops and densely populated areas, it is reasonable to suppose that there are many connected devices emitting numerous Wi-Fi signals. That being said, note that an interference phenomenon is also sometimes reported in the literature, advocating for reduced performance of Wi-Fi sensors when the surrounding crowd is high \citep{quteprints62727,franssensImpactMultipleInquires2010}.

Thirdly, specific stops appear to concentrate (negative) errors, hinting at underestimates of the boarding/alighting volumes at these stops with the \emph{Laflowbox} technology.

\subsection{Inference of the O-D error on the basis of the boarding and alighting errors}

For developers of the Wi-Fi based technology and for transport analysts, it would be highly beneficial to know the accuracy of the O-D matrices estimated on this basis. Of course, this cannot rely on day-to-day exhaustive O-D surveys (which would anyway make the deployment of the Wi-Fi sensors pointless). By contrast, boarding and alighting volumes are more easily accessible (e.g., using optical counts as in the present case). Therefore, one would like to infer the O-D error $Err$ on the basis of the boarding and alighting errors $Err^{(in)}$ and $Err^{(out)}$ (see Sec.~\ref{sub:methods_notations}), for instance by being able to predict their ratios, i.e., the error ratios $err^{(in)}$ and $err^{(out)}$. As we will see, the observations made in the previous paragraphs, although enlightening, are insufficient to capture these ratios.

\subsection{Modeling the O-D errors with simple models of random noise}

\begin{table}[ht] 
\footnotesize 
\begin{center} 
\caption{Summary of the simple noise models used to reflect the O-D estimation error $\delta T_{io}$ in this study. $\xi_{io}$, $\xi'_{io}$, $\xi_{i}$ and $\xi_{o}$ are all i.i.d. random variables uniformly distributed in $[-\sigma,\sigma]$, with $\sigma>0$; the $Z[\xi]$ are renormalization terms merely aimed at conserving the sum of shares equal to 1. Refer to the defining equations for the other notations.} 
\end{center}
\begin{tabular}{p{4.5cm}p{5cm}m{3cm}}
\hline
&  $\delta T_{io}= - Z[\xi] + \dots$
& \hfil Definition 
\vspace{0.3cm}
\\
\hline 
\textbf{Additive noise}
& \hspace{1cm} $\dots +  \xi_{io}  $
 & \hfil Eq.~\ref{eq:add}
\\ 
\textbf{Multiplicative noise}
& \hspace{1cm} $\dots +  \xi_{io}\, T_{io} $
& \hfil Eq.~\ref{eq:mult}
\vspace{0.3cm}
\\ 
\textbf{Combined additive and multiplicative noises} 
& \hspace{1cm} $\dots +  \xi_{io} + \xi^{\prime}_{io}\, T_{io} $
& \hfil Eq.~\ref{eq:add-mult}
\vspace{0.3cm}
\\ 
\textbf{Noise on short O-D trips} 
& \hspace{1cm} $\dots +  \xi_{io} \, \Theta(2-|i-o|)  $
& \hfil Eq.~\ref{eq:short}
\vspace{0.3cm}
\\ 
\textbf{Noise on central O-D stops} 
& \hspace{1cm} $\dots + \xi_{io} \cdot  \mathbf{1}_c(i) \cdot  \mathbf{1}_c(o) $
& \hfil Eq.~\ref{eq:central}
\vspace{0.3cm}
\\ 
\textbf{Uncertainty on the boarding stop} 
& \hspace{1cm} $\dots +  \xi_{i}  $
& \hfil Eq.~\ref{eq:in-noise}
\vspace{0.3cm}
\\ 
\textbf{Uncertainty on the alighting stop} 
& \hspace{1cm} $\dots + \xi_{o}$
& \hfil Eq.~\ref{eq:out-noise}
\vspace{0.3cm}
\\ 
\hline 
\end{tabular} 

\label{table2} 
\end{table}

\subsubsection*{Additive noise}

With the foregoing goal in mind, let us start by describing the structure of the estimation error $\delta T_{io}$ with  the simplest possible random noise models, namely the uncorrelated additive noise defined in Eq.~\ref{eq:add}. Of course, by playing with the amplitude $\sigma$ of this noise, one can trivially reproduce the RMS O-D error $Err$.

\begin{figure}[ht]\vspace*{4pt}
\centerline{\includegraphics[height=8cm]{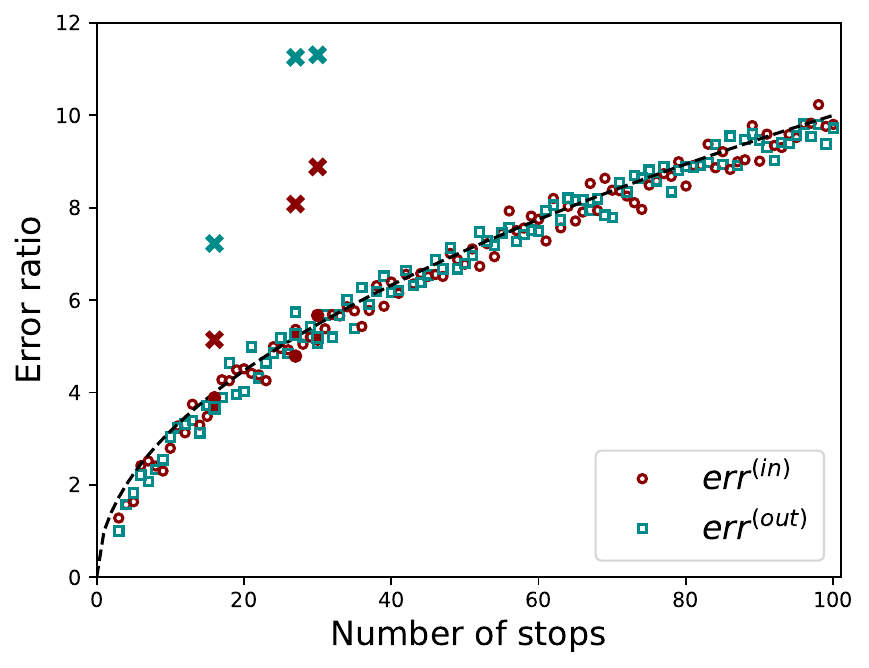}}
\caption{ Ratios of the boarding (red) or alighting (blue) errors over the error on O-D shares. Simulations with an additive noise operating on a synthetic O-D matrix are represented by empty circles and squares, while full symbols indicate the action of this additive noise model on the genuine reference O-D matrices. For each number of stops $N$, the results have been averaged over 10 realizations of the noise. The empirical data (O-D survey and Wi-Fi data) for lines T1 (15 stops), T2 (30 stops) and T3 (27 stops) are shown as large crosses. The law  $err=\frac{1}{\sqrt{N}}$ predicted by Eq.~\ref{eq:central_lim_theo} is represented as a dashed line.}
\label{fig5}
\end{figure}

But its relation with $Err^{(in)}$ and $Err^{(out)}$ cannot be replicated in this way, let alone the structure of the error of Figure~\ref{fig12}. To show this, we begin by considering the synthetic O-D-like matrices we constructed for any number of stops $N$ (the random uniform matrices defined in Sec.~\ref{sub:synth-OD}) and applying additive noise $\xi_{io}^{\mathrm{(add)}}$ to their entries $T_{io}$. We plot the boarding and alighting error ratios as empty circles and squares in Figure~\ref{fig5}. These error ratios nicely follow a $\sqrt{N}$ law. In other words, the mean absolute errors $Err^{(in)}$ and $Err^{(out)}$ on  the boarding/alighting volumes  are 
 greater by only a factor  $\sqrt{N}$ than the mean error $Err$ on O-D share (whereas the volumes are larger than the individual O-D share  by a full $N$ factor, on average). This result is not connected to the structure of the synthetic O-D matrix: Applying the same additive noise $\xi_{io}^{\mathrm{(add)}}$ to the entries of the reference O-D matrix yields identical results (full symbols in Figure~\ref{fig5}). This scaling in the presence of additive noise is easy to grasp; starting from Eq.~\ref{eq:Err_in}, one calculates

\begin{linenomath}
\begin{eqnarray}
N\cdot\Big[Err^{(in)}\Big]^2 &=& \sum_{i=1}^{N}(\delta T_{i\bullet})^2 \nonumber \\
&=& \sum_{i=1}^{N}\Big(\sum_{o=1}^{N}\delta T_{io}\Big)^2 \nonumber \\
& \approx &\sum_{i=1}^{N}\sum_{o=1}^{N}(\delta T_{io})^2
\end{eqnarray}
\end{linenomath}
In the last line, cross-terms of the form $\delta T_{io}\cdot \delta T_{i'o'}=\xi_{io} \cdot \xi_{i'o'} - Z^{\mathrm{(add)}}[\xi] \cdot (\xi_{io} + \xi_{i'o'}) + Z^{\mathrm{(add)}}[\xi]^2$, where $i \neq i'$ or $o \neq o'$, have been neglected, because these random uncorrelated contributions will be sometimes positive and sometimes negative; they will cancel out on average and/or when $N$ gets very large. Therefore, after recognizing the definition of $Err$ (Eq.~\ref{eq:Err}) in the last line, one finally arrives at the observed law

\begin{equation}
    Err^{(in)} \approx \sqrt{N}\,Err.
    \label{eq:central_lim_theo}
\end{equation}

Unfortunately, the \emph{empirical} error ratios, marked as crosses in Figure~\ref{fig5}, significantly deviate from these predictions for all bus lines. For instance, on the T3 line ($N = 27$), we observe
\[Err^{(in)} = 0.018;\ Err^{(out)} = 0.026;\ Err = 0.0023\] 
\[err^{(in)}= 8.08; err^{(out)} = 11.25\] 
Instead, Eq.~\ref{eq:central_lim_theo} would predict 
\[err^{(in)} = err^{(out)} \simeq \sqrt{N}\text{ with }\sqrt{N} \simeq 5.2\ (N = 27). \]
Not only are the empirical error ratios larger than predicted by the additive noise model, but in addition, quite unexpectedly, they differ significantly between boarding and alighting (the latter is much higher).
These discrepancies point to singular characteristics in the detection error structure that originates from the Wi-Fi-based O-D assignment; additive noise applied to the O-D matrix entries fails to capture these.

\subsubsection*{Multiplicative noise}
Turning to multiplicative noise (Eq.~\ref{eq:mult}) or to a combination of additive and multiplicative noises (Eq.~\ref{eq:add-mult}) does not remedy the foregoing issue. Nor does the recourse to clamped or unclamped random noises (see \emph{Methods}). The boarding or alighting error ratios simulated under these noise assumptions still follow a $\sqrt{N}$-law (see Fig.~\ref{fig:SM_A} for graphical results). These noises fail to capture the experimentally observed relation between O-D error on the boarding or alighting errors.

\subsubsection*{Noise centered on short and central stops}
\begin{figure}[ht]\vspace*{4pt}
\centering
\includegraphics[width=0.45\textwidth]{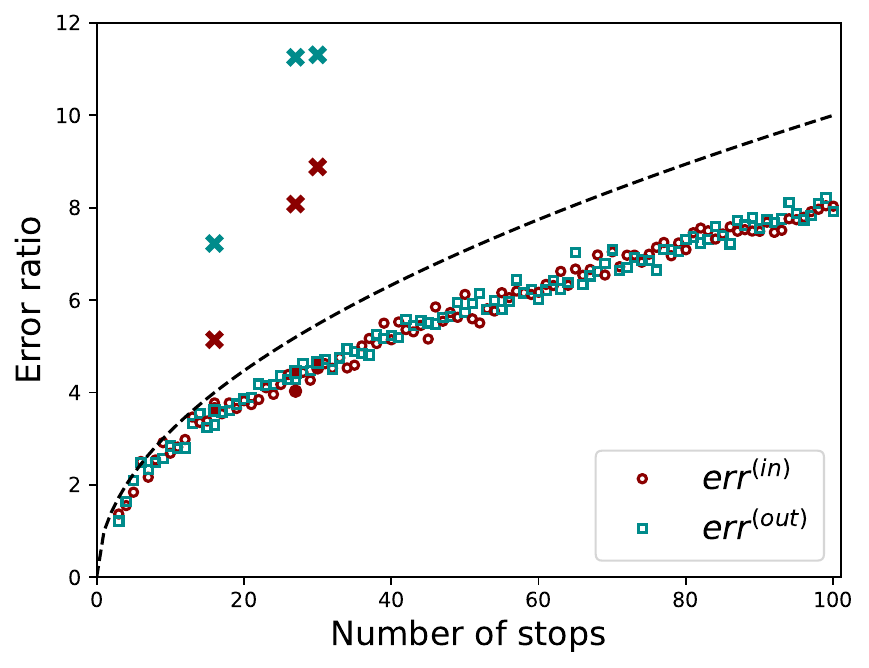}~\includegraphics[width=0.45\textwidth]{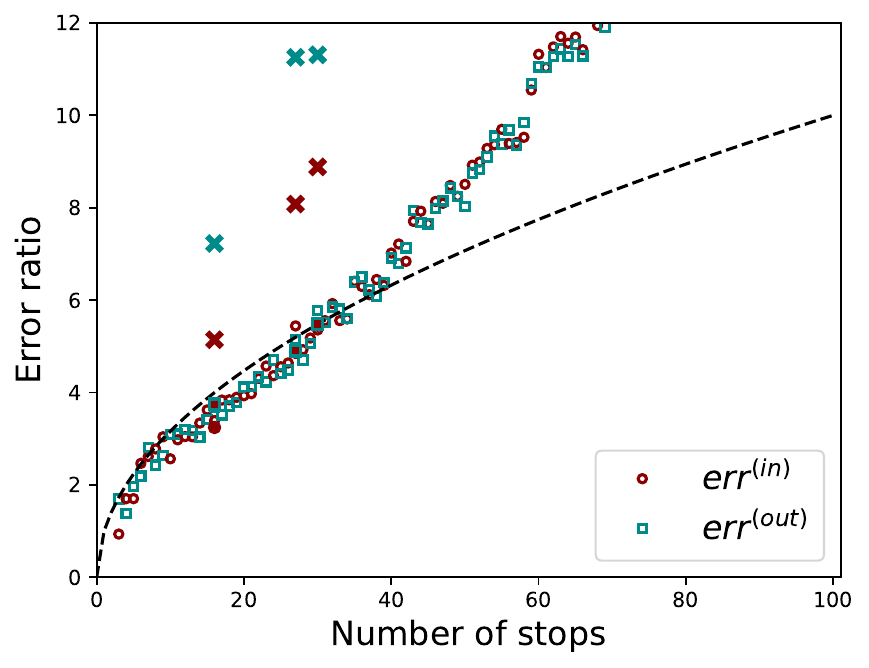}
\caption{Boarding (red) and alighting (blue) error ratios, $err^{(in)}$ and $err^{(out)}$, obtained for stochastic noises concentrated on short O-D (left) or short O-D and central ones (right). The noise amplitude $\sigma$ was set to 0.1. Refer to Figure~\ref{fig5} for the rest of the caption. }
\label{fig8}
\end{figure}

The foregoing noise models affect the  the O-D matrix uniformly. Given their failure, we turn to heterogeneous noise. In light of our direct observations about the structure of the O-D error (Sec.~\ref{sub:obs}), we consider the introduction of noise that specifically targets short O-D routes (no more than two stops apart) or central stops, as proposed in Eq.~\ref{eq:short} and Eq.~\ref{eq:central}. These noise structures produce error matrices that are less uniform and look more similar to those of Figure~\ref{fig12}. Moreover, they lead to \emph{boarding} and \emph{alighting} error ratios that deviate from the $\sqrt{N}$ law found so far, as shown in Figure~\ref{fig8}; in particular, the noise that concentrates on central O-D can yield error ratios somewhat closer to the empirical ones (right panel of Figure~\ref{fig8}). Here, using clamped noises had a marked effect, while making the noise on short O-D only positive (to get an O-D error matrices that look more similar to Figure~\ref{fig12})) had no incidence on the error ratios.

However, experimentally, there is always a difference between the ratio $err^{(in)}$ and the ratio $err^{(out)}$, which the model fails to reproduce: they fall short of capturing the boarding error ratio and the alighting error ratio at the same time.  Combining these heterogeneous noise models with a residual additive noise on the rest of the O-D matrix, to account for the fact that short O-D and central stops are not the only sources of error, was not fruitful either. Once again, applying these noise models to the reference O-D matrices produced error ratios comparable to those obtained with our synthetic O-D matrices.

\subsubsection*{Asymmetric uncertainties between boarding and alighting}
Therefore, we are led to consider the asymmetry between boarding and alighting with respect to the Wi-Fi sensor operations.
Given that passengers are likely to wait a few minutes at the bus stop before boarding, some connected objects are likely to be detected as soon as the bus stop enters the antenna's field of detection. On the contrary, it is likely that passengers who get off the bus do not linger as long as boarding passengers at the stop (unless the passenger is waiting for a transfer, but this remains marginal in our experience, with a transfer rate of 1.17 in 2019 \citep{OMMeR} on the whole network of Rouen). Their connected objects therefore quickly become undetectable, and the antenna of the sensor will not receive many signals in the surroundings of the bus stop. As connected objects do not emit regularly, it is also possible for a connected object to leave the bus without having transmitted a signal for some time (one or two stops previously), making the estimation of the alighting stop more difficult. Another contributor to the uncertainty about the alighting stop is the randomization of the MAC addresses that is operated once in a while by a growing number of smartphones, which ends a signal detection before the passenger actually alights -- but it may also artificially create a new trip if the MAC changes close enough to a boarding stop. \emph{If no such new trip is generated}, a difference appears in the errors made in relation to boarding vs. to alighting volumes, which should be lower for the former.

\begin{figure}[ht]\vspace*{4pt}
\centering
\includegraphics[height=8cm]{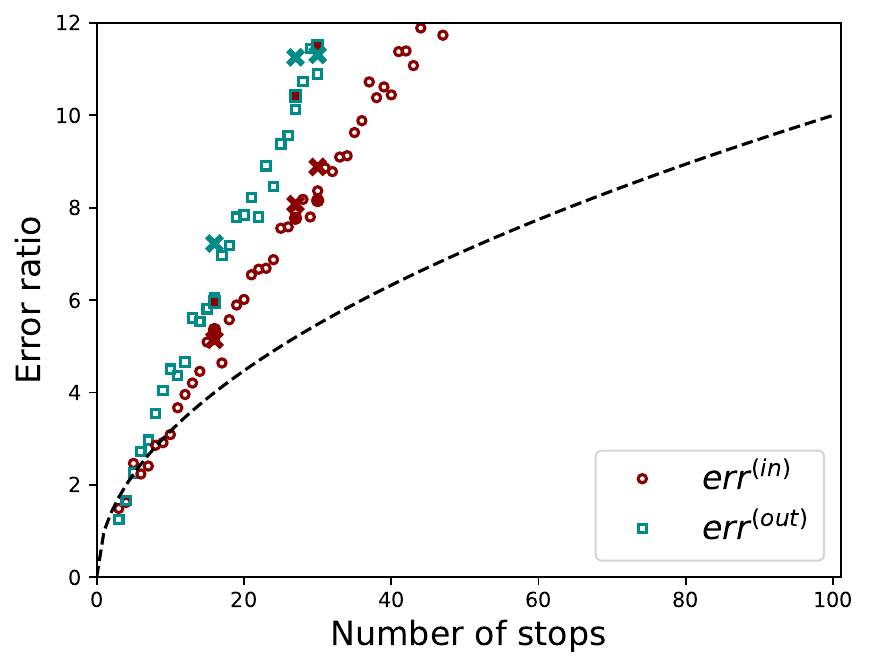}
\caption{Boarding (red) and alighting (blue) error ratios, $err^{(in)}$ and $err^{(out)}$, obtained by introducing an asymmetric uncertainty between boarding and alighting. The noise model combines an uncertainty associated with the boarding stop (of amplitude $\sigma=0.03$), an uncertainty associated with the alighting stop (of amplitude $\sigma=0.045$), and a homogeneous additive noise (of amplitude $\sigma=0.1$). Refer to Figure~\ref{fig5} for the rest of the caption.}
\label{fig9}
\end{figure}

Thus, we introduce distinct uncertainties with respect to boarding \emph{vs.} alighting, materialized by the boarding and alighting noises of Eq.~\ref{eq:in-noise} and Eq.~\ref{eq:out-noise}. In matricial format, these noises operate on row or columns of the O-D matrix. The results in terms of error ratios are shown on Figure~\ref{fig9}, for a noise amplitude $\sigma=0.03$ for boarding and $\sigma=0.045$ for alighting, together with an additive noise of amplitude $\sigma=0.1$.  
Clearly, this model succeeds in discriminating the error ratio associated with alighting and that associated with boarding, and in capturing their empirical values. 

\subsection{Practical applications}

The foregoing results point to conditions on the noise such that the ratios $err^{(in)}$ and $err^{(out)}$ match the experimental ones. We now hint at practical applications of this result.

\begin{figure}[ht]\vspace*{4pt}
\centering
\includegraphics[height=8cm]{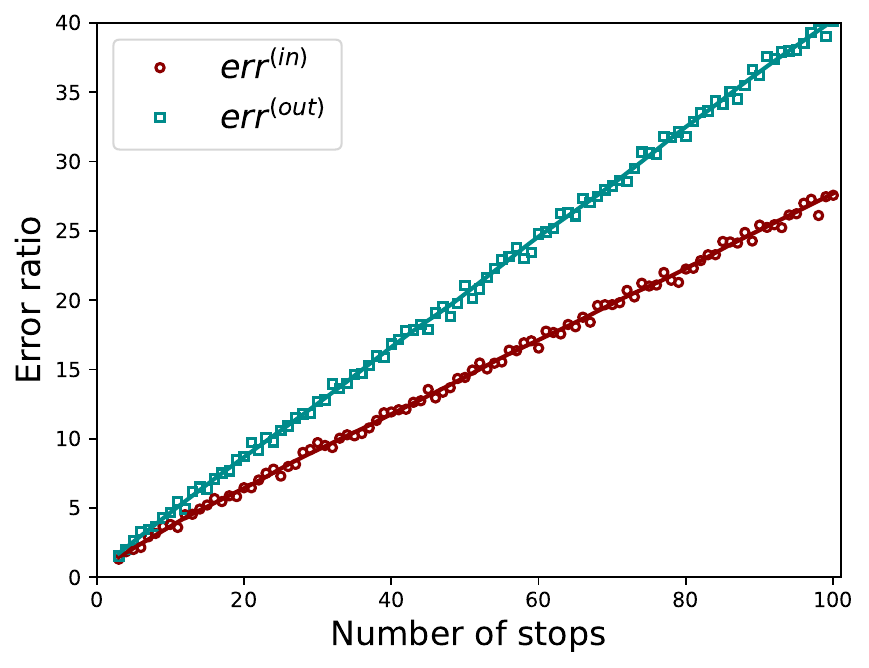}
\caption{Lowess regression for O-D errors/Boarding shares errors (red) and  for O-D errors/Alighting shares errors (blue).}
\label{fig11}
\end{figure}

Knowing the number of stops, the value of this ratio can be computed, so that it is possible to estimate the error on O-D shares knowing only $Err^{(in)}$ or $Err^{(out)}$ (which, as a reminder, are readily available if an optical counter is installed). 
To illustrate this concretely, we use a locally weighted scatter plot smoothing (Lowess) which is a non-parametric regression. That gives the smoothed curved adjusted to our points of abscissa the number of stops in the matrix and of ordinate the ratio $err^{(in)}$ or $err^{(out)}$. Using a degree of smoothing of ${20\%}$ the regression lines presented in Figure~\ref{fig11} are obtained. As an example, using this approach $err^{(in)}\simeq 7$ for $N=22$ which will allow us to get $Err$ knowing $Err^{(in)}$ for a bus line with 22 stops.

\begin{figure}[ht]\vspace*{4pt}
\centering
\includegraphics[height=8cm]{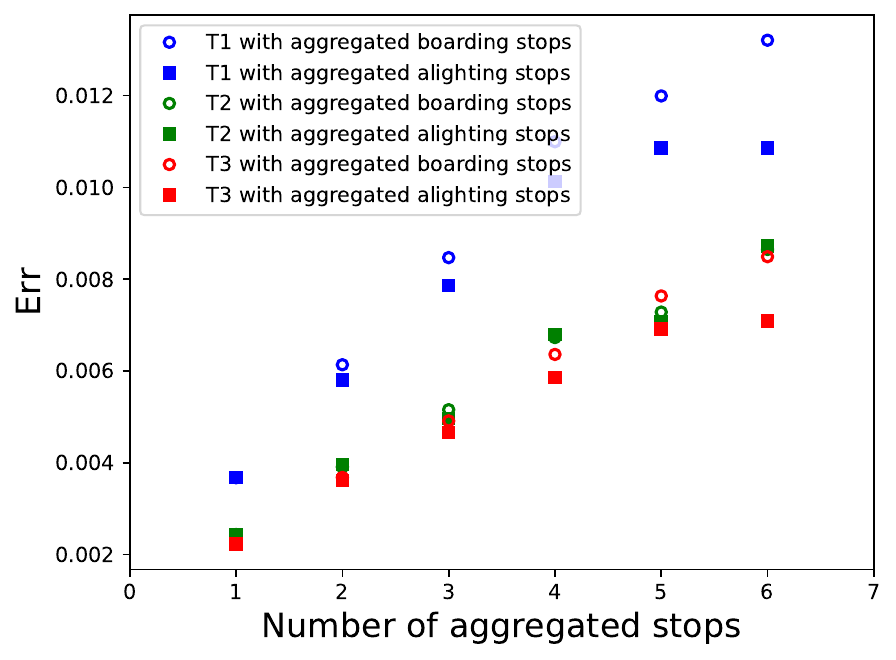}
\caption{O-D error obtained upon aggregation of either alighting stops or boarding stops (see the legend), depending on the number of successive spots that are bundled together.}
\label{figAggr}
\end{figure}

Besides, the insight obtained into the main sources of error hints at strategies to aggregate the O-D share estimates in a way that lessens their inaccuracy. Suppose that one is not interested in the exact boarding or alighting stop, but in an approximate area. Then, stops may be grouped into aggregates of $n$ stops, the first aggregate gathering stops from 0 to $n-1$, the second one gathering stops from $n$ to $2n-1$, etc. Now, the asymmetry that we unveiled between boarding and alighting uncertainties implies that we had better aggregate alighting stops than boarding stops, because the uncertainty about alighting is somewhat larger. This is demonstrated in Figure~\ref{figAggr}, where the empirical error on O-D shares is calculated under the assumption that different numbers $n$ of either boarding or alighting stops have been aggregated. (Since $Err$ is the root-mean square error \emph{per matrix entry}, $Err$ is expected to increase with the aggregation number $n$, because the values of the entries then get larger.) The difference between boarding and alighting aggregations is not huge, but it is clearly visible (with the exception of line T2). 

\section{Conclusion and future research}
\label{sec:conclusion}

Collecting mobility data is essential to understand the travel behavior of inhabitants and to ensure the sustainable development of transport infrastructures. In recent years, following the increasing cost of travel surveys and the development of technologies, new data sources are emerging. Among them, Wi-Fi sensors installed in city buses allow real and continuous collection of mobility data. This passive method requires no effort from passengers and represents a marginal cost for the public authority. Some previous works have compared O-D matrices generated with Wi-Fi and those derived from ground truth data, and results seem promising. However, the question of data quality remains. Moreover, O-D surveys are conducted only every five or ten years and cannot be used regularly to assess the quality of O-D matrices built from Wi-Fi data. The paper addresses this problem by proposing a method for estimating the error on the O-D pairs using the error on the boarding and alighting passengers' shares, the latter being readily and continuously available in all public transportation systems where optical counting devices are installed.

By applying different noise structures to a stylized (random) O-D matrix to account for the O-D assignment error, different relations between the error on the O-D pairs shares and the error on the boarding (or alighting) passengers shares are obtained. Interestingly, simple noise structures are not compatible with the empirical error ratios, notably its asymmetry between boarding and alighting. Instead, based on the hypothesized origin of the error, a relevant noise structure that reproduces the empirical observations is identified. This is not a mere calculation of the error, as in the previously cited works, but a proper modelling of the error.

Our key observations regarding the structure of the estimation error are thus

\begin{itemize}
\item The error made on Wi-Fi O-D estimates compared to the one from O-D survey is greater on short O-Ds.
\item This error is also greater on central stops. 
\item The error made on estimates of the number of alighting passengers with Wi-Fi data is greater than that made on estimates of the number of boarding passengers.
\end{itemize}

This paper makes both theoretical and operational contributions. From a theoretical point of view, the proposed method leads to a relationship between the error on O-D pairs shares and the error on boarding (or alighting) passengers shares. This way, it is possible to quantify the error on a continuous basis, thus complementing recent work enabling a one-time comparison. The results highlight the rather complex relationship between the error on O-D pairs shares and the error on boarding or alighting passengers shares, and the difference in the structure of the error between boarding and alighting passenger shares. In the end, it is possible to generate, using Wi-Fi data, an O-D matrix associated with an error structure. 

From an operational point of view, this work is of particular interest to transport practitioners, who can thus learn about the error structure affecting the O-D matrix generated from passive data. At any given time, access to Wi-Fi data and optical counts (the number of boarding and alighting passengers) is all that is needed to assess to what extent the O-D matrix deviates from the actual mobility behaviors. It is then possible to gauge the relevance of Wi-Fi data on a continuous basis without needing O-D surveys, by nature expensive, time-consuming and infrequent. Moreover, the methodology can be used by analysts to assess the reliability of the devices increasingly used in public transport to produce mobility data from Wi-Fi data. Finally, the error in the Wi-Fi O-D matrix obtained from this work could help to improve the Wi-Fi data collectors. Indeed, since the main errors occur in central zones where many noisy signals are detected, this suggests that using a shorter range antenna could improve Wi-Fi data collection. This is also consistent with the literature, since \cite{algomaiah2022utilizing} found optimal detection radii shorter than those of the Wi-Fi sensors used in the present study.

As further works, it would be of particular interest to use this model to withdraw the estimated noise from Wi-Fi O-D estimates. This would allow having a picture of trips on the network, as close to the reality as possible. Also, the work could be replicated on another network to see, first, if the modeling is easily replicable and also to see if the same relations between the errors apply. A comparison between the modeled error on different networks could reveal interesting insights on travel behaviors of each network. It could also present differences on the structure of the error in Wi-Fi O-D estimates of each network. That will lead to interesting speculations on the proximity of O-D surveys with reality. 



\section{CrediT authorship contribution statement and Declaration of Competing Interest}
\label{sec:creditAuthor}
LF: Conceptualization, Methodology, Data curation, Formal analysis, Visualization, Validation, Writing – original draft. 

CB: Conceptualization, Methodology, Supervision, Validation, Writing – review \& editing. 

AN: Conceptualization, Methodology, Visualization, Validation, Writing – review \& editing. 

PB: Conceptualization, Supervision. 

All the authors confirm that this work is original and has not been published elsewhere. It is not currently under consideration for publication elsewhere.

The authors declare that they have no known competing financial interests or personal relationships that could have appeared to influence the work reported in this paper.

\section{Acknowledgements}
\label{sec:acknowledgements}
This research was conducted as part of a research agreement between EXPLAIN, the Urban Planning, Economics and Transport Laboratory (LAET) and the Actuarial and Financial Sciences Laboratory (LSAF). The authors acknowledge them for the financial support. The Métropole Rouen Normandie is also thanked for providing the data.

\clearpage

\appendix
\renewcommand\thefigure{S\arabic{figure}}    
\setcounter{figure}{0} 

\section{Additional figures}
This Appendix gathers additional figures to support the findings reported in the main text.

\begin{figure}[ht]
\centering
\includegraphics[draft=False,height=5cm]{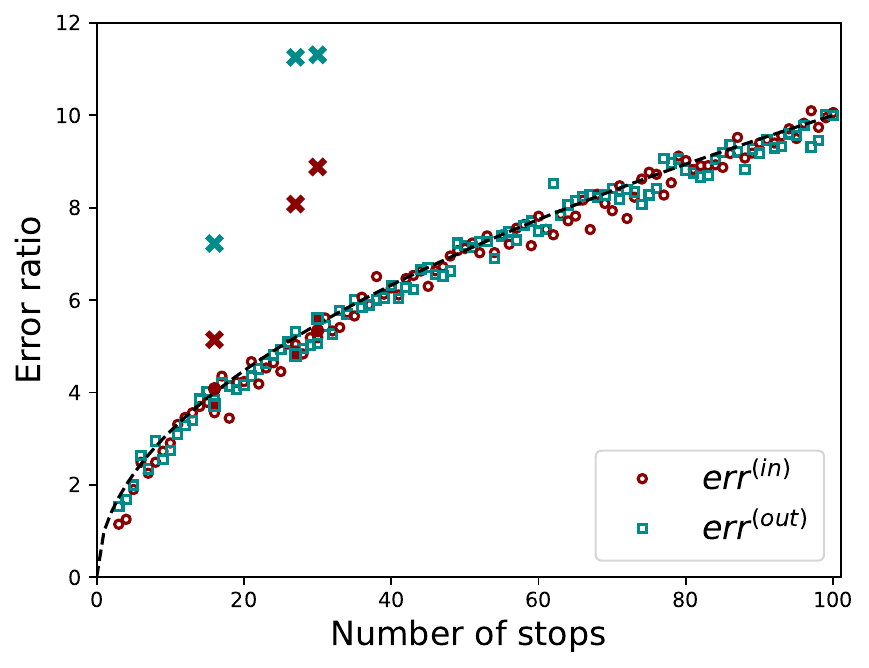}\hspace*{5mm}\includegraphics[draft=False,height=5cm]{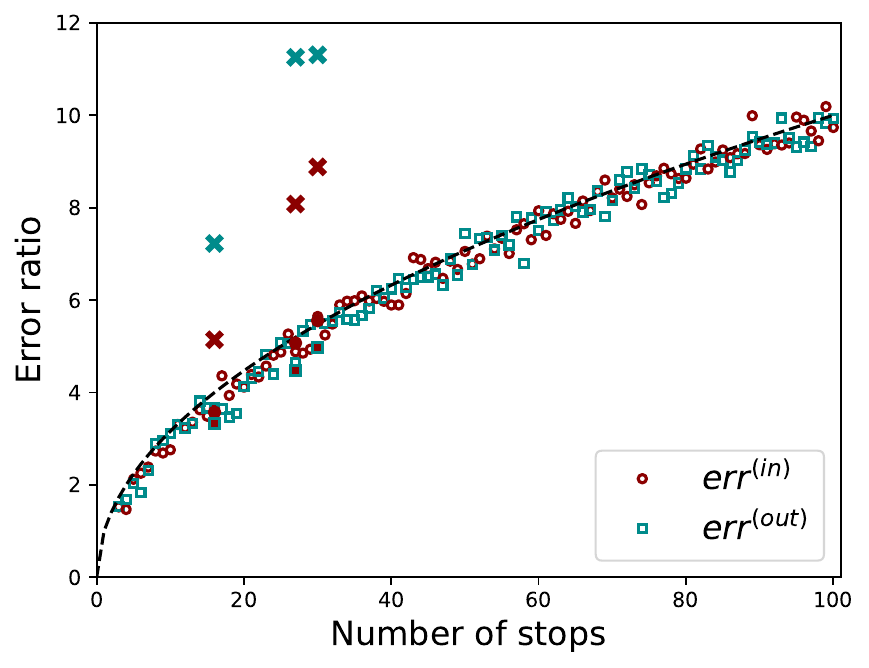} \\
\includegraphics[draft=False,height=5cm]{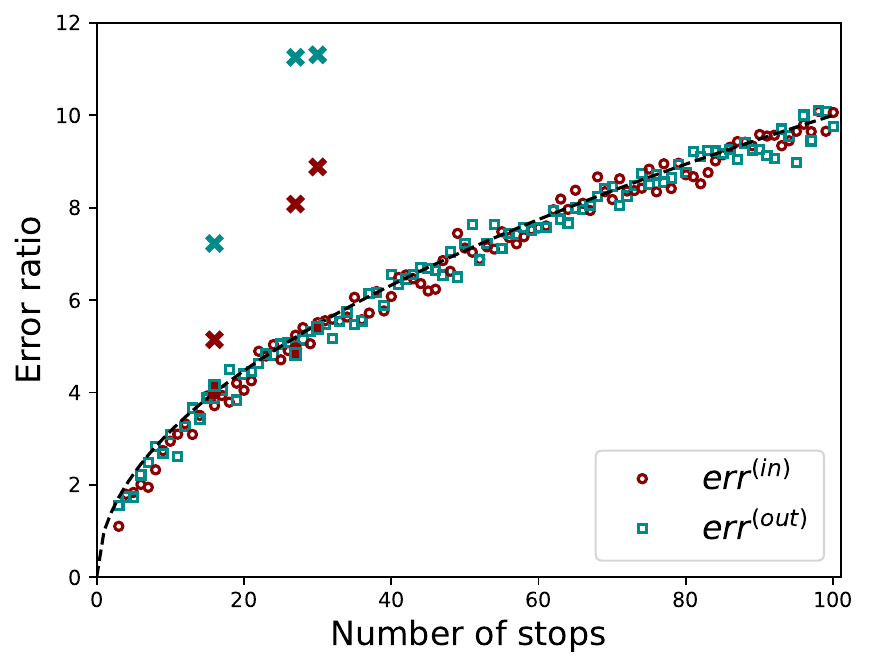}
\caption{Ratio of the errors on O-D errors and on Boarding shares (red) or Alighting shares errors (blue). Modeling of the error with (a) additive noise; (b) multiplicative noise; (c) a combination of both noises.
\label{fig:SM_A}}
\end{figure}

\clearpage






\end{document}